\documentclass[
reprint,
twocolumn,
amsmath,amssymb,
superscriptaddress,aps,
pre
]
{revtex4-2}
\usepackage[utf8]{inputenc}
\usepackage{graphicx}
\usepackage{hyperref}
\usepackage{url}
\usepackage[normalem]{ulem}

\usepackage{xcolor}

\usepackage{mathtools}

\begin{document}

\title{Spontaneous symmetry breaking in ride-sharing adoption dynamics}

\author{Henrik Wolf}
\affiliation{Chair for Network Dynamics, Institute for Theoretical Physics and Center for Advancing Electronics Dresden (cfaed), Technische Universität Dresden, 01062 Dresden, Germany}

\author{David-Maximilian Storch}
\affiliation{Chair for Network Dynamics, Institute for Theoretical Physics and Center for Advancing Electronics Dresden (cfaed), Technische Universität Dresden, 01062 Dresden, Germany}

\author{Marc Timme}
\affiliation{Chair for Network Dynamics, Institute for Theoretical Physics and Center for Advancing Electronics Dresden (cfaed), Technische Universität Dresden, 01062 Dresden, Germany}

\author{Malte Schröder}
\affiliation{Chair for Network Dynamics, Institute for Theoretical Physics and Center for Advancing Electronics Dresden (cfaed), Technische Universität Dresden, 01062 Dresden, Germany}

\begin{abstract}
Symmetry breaking ubiquitously occurs across complex systems, from phase transition in statistical physics to self-organized lane formation in pedestrian dynamics.
Here, we uncover spontaneous symmetry breaking in a simple model of ride-sharing adoption. We analyze how collective interactions among ride-sharing users to avoid detours in shared rides give rise to spontaneous symmetry breaking and pattern formation in the adoption dynamics. These dynamics result in bistability of high homogeneous and partial heterogeneous adoption states, potentially limiting the population-wide adoption of ride-sharing. Our results provide a framework to understand real-world adoption patterns of ride-sharing in complex urban settings and support the (re)design of ride-sharing services and incentives for sustainable shared mobility.
\end{abstract}

\maketitle

\section{Introduction}
Spontaneous symmetry breaking and related pattern formation dynamics are ubiquitous across complex systems. Examples range from phase transition and phase separation in standard many-particle systems in statistical physics \cite{sethna2006statistical, Coleman2015, Sander2009} over the emergence of non-trivial structures in complex networks \cite{Strogatz2001, Reka2002, Newman2003} to the dynamics of human mobility systems \cite{barbosa2018_mobility_models, Caldarelli2018}. Often, these intriguing collective dynamics emerge from apparently simple interactions \cite{Holovatch2017}. In human mobility, for example, symmetry breaking occurs in the emergence of congestion phenomena in car traffic \cite{Schreckenberg92, Helbing2001, Helbing2002, Helbing2004}, in trail and lane formation in self-organized pedestrian movement \cite{Helbing1997b, helbing2000freezing}, or in the evolution of cities and large-scale commuting patterns \cite{Gonzalez2008, Song2010b, Mazzoli2019}.

In recent years, the complexity of human mobility has further increased. On-demand ride-hailing and ride-sharing services (also referred to as ride-pooling) have grown to be a major part of urban mobility \cite{furuhata2013ridesharing}. Like with conventional taxi services, users request a ride to a desired destination and are then picked up and delivered with door-to-door service. However, unlike with standard taxi rides, users of ride-sharing services may be matched into a shared ride with other users who will be delivered to their respective destinations in the same vehicle. The combination of multiple trips into one vehicle potentially reduces the total distance driven compared to individual mobility options which makes ride-sharing a promising alternative to reduce congestion and emissions with growing urban mobility demands \cite{santi2014_shareabilityNetworks, vazifeh2018_minimumFleetProblem, Anair2020, Jenn2020}. 
Recent analyses have been focused on the collective dynamics of the vehicle fleets and drivers, the emerging universal properties of the efficiency of ride-sharing across different settings \cite{santi2014_shareabilityNetworks, tachet2017scaling, Molkenthin2020}, and potential negative socio-economic consequences of currently prevailing ride-hailing services \cite{Schroeder2020, Erhardt2019}. Yet, it remains an open problem to understand the collective dynamics from the user perspective with particular focus on how these dynamics promote or inhibit the adoption of ride-sharing \cite{Storch2020, Alonso-Gonzalez2020, Ruijter2020}.

Here, we reveal spontaneous symmetry breaking in the adoption dynamics of ride-sharing users. Based on a recently proposed game-theoretic model of ride-sharing adoption dynamics \cite{Storch2020}, we demonstrate how incentives to avoid detours in shared rides give rise to strategic interactions between the users. Long-range inhibition of ride-sharing adoption intrisic to these interactions results in stable heterogeneous patterns of ride-sharing adoption that break the symmetry of the interactions between two users. These adoption patterns coexist with spatially homogeneous states of high ride-sharing adoption and potentially limit the aggregate adoption of the ride-sharing service at the population level. By combining exact \mbox{(semi-)}analytical evaluations with detailed numerical simulations, we reveal how the emerging patterns change with the total demand and why ride-sharing services may not be efficient despite a high number of potential users. Our results may help to better predict ride-sharing adoption and potential inefficiencies in complex urban environments, contributing to a better understanding of the interaction between ride-sharing and other modes of (public) transportation.

\section{Model}

\begin{figure*}[ht]
    \centering
	\includegraphics{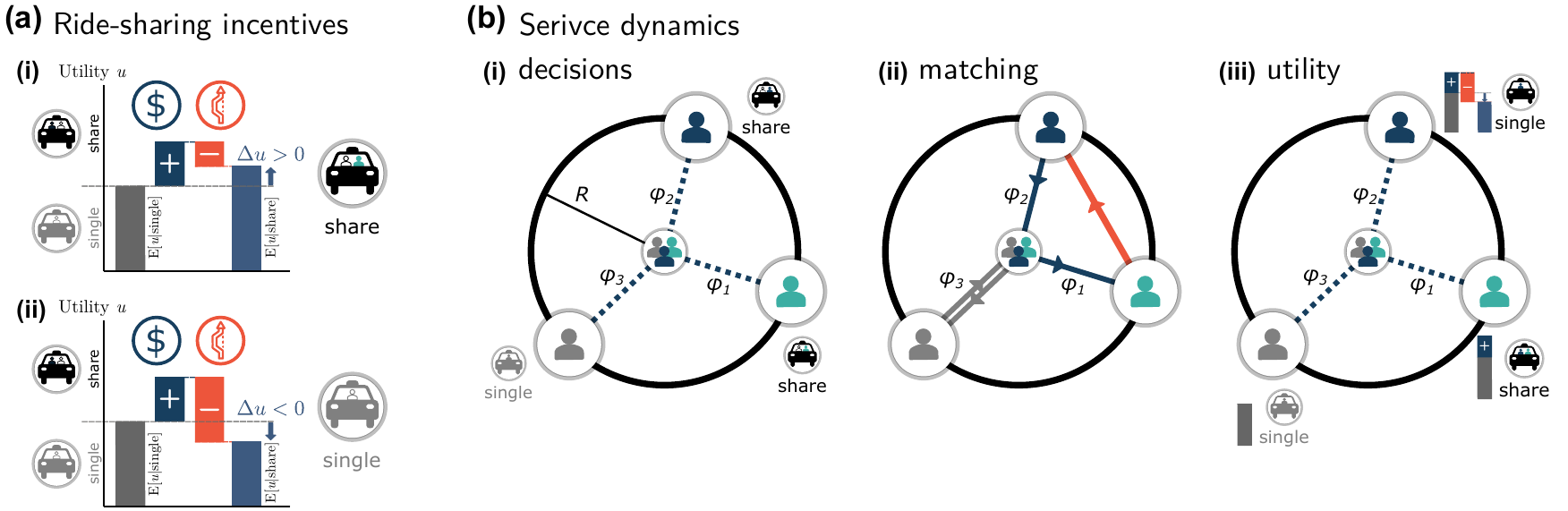}
    \caption{
        \textbf{Incentives and interactions in ride-sharing adoption dynamics.} 
        (a) Incentives govern the adoption of ride-sharing services (adapted from \cite{Storch2020}). Financial incentives $a\,d_\mathrm{req}$ proportional to the requested trip distance $d_\mathrm{req}$ incentivize sharing (dark blue). Potential detours $b\,d_\mathrm{det}$ from delivering other passengers first in a joint ride disincentivize sharing (orange). (i) Users prefer shared rides if the utility difference $\Delta u = a\,d_\mathrm{req} - b\,d_\mathrm{det}$ compared to single rides is positive [Eq.~\eqref{eq:expected_utility}]. (ii) If the utility difference is negative due to long detours, they prefer single rides. 
        (b) In a simplified one-to-many demand setting with a single origin in the center and destinations $\varphi$ uniformly independently distributed on a circle with radius $R$, (i) users request shared or single rides according to their ride-sharing adoption probability $p(\varphi)$ which depends on their destination $\varphi$. (ii) They are matched with other users into shared rides to minimize the total distance driven, including the return trip. (iii) Users realize their utility for the (shared) ride depending on their detour [orange, Eq.~\eqref{eq:detour}]. Over time, the users learn the expected utility difference $E[\Delta u(\varphi)]$ between single and shared rides for their destination $\varphi$ and update their ride-sharing adoption probability [Eq.~\eqref{eq:replicator_dynamics}].
    }
    \label{fig:FIG1_model}
\end{figure*}

\subsection{Ride-sharing incentives}
The decision of a ride-hailing user to request a single or a shared ride depends on the possible benefits (or utilities) of the two options \cite{Storch2020, Alonso-Gonzalez2020, Morris2019, Sarriera2017, LO2018,Schwieterman2019} (see Fig.~\ref{fig:FIG1_model}a). Users balance two competing incentives: (i) Service providers incentivize shared rides through financial discounts compared to single rides. These discounts are often percentage discounts of the single trip fare, proportional to the direct trip distance $d_\mathrm{req}$ of the requested trip, and are granted independently of whether the ride is actually \textit{matched} with another user. (ii) Potential detours $d_\mathrm{det}$ from successfully matched rides discourage requesting shared rides. Detours mediate repulsive strategic interactions between the users' decisions and underlie the collective phenomena observed in the adoption dynamics. 

We take the utilities to be proportional to the relevant distances. The positive utility $a\,d_\mathrm{req}$ of the financial discount for a shared ride request to destination $\varphi$ is proportional to the direct trip distance $d_\mathrm{req}$ and the negative incentives $b\,d_\mathrm{det}$ are proportional to the detour $d_\mathrm{det}$ of the shared ride. The proportionality factors $a$ and $b$ describe the importance of financial incentives and detour, respectively. The financial discount is constant since the direct trip distance $d_\mathrm{req}(\varphi)$ depends only on the destination $\varphi$. In contrast, the (expected) detour, and thereby also the overall utility $u$, depend on the destinations and sharing decisions of other users that jointly affect which requests are matched with each other. 

Combining both the financial and the detour incentives, the decision to book a shared or a single ride depends only on the expected utility difference $E\left[\Delta u(\varphi)\right]$ for a user with destination $\varphi$. This difference measures the expected incremental utility of booking a discounted shared ride with a potential detour compared to the constant utility of booking a direct single ride,
\begin{eqnarray}
    E\left[\Delta u(\varphi)\right] &=& E\left[u\left(\varphi\right)\,\middle|\,\mathrm{share}\right] - E\left[u\left(\varphi\right)\,\middle|\,\mathrm{single}\right]\nonumber\\
    &=& a\,d_\mathrm{req}(\varphi) - b\,E\left[d_\mathrm{det}(\varphi)\,\middle|\,\mathrm{share}\right] \,, \label{eq:expected_utility}
\end{eqnarray}
where $E\left[X \mid s\right]$ denotes the expectation value of the random variable $X$ with respect to the destinations and sharing decisions of all other users conditional on the user's own sharing decision $s$. Over time, users learn the expected utility difference between requesting a shared or a single ride for a given trip and update their decisions accordingly [see Eq.~\eqref{eq:replicator_dynamics} below]. 

\subsection{Ride-sharing model}
Ride-sharing in its most basic setting combines several concurrent trip requests with a common origin $o$, for example from high-demand locations such as train stations or airports, to similar destinations $\varphi_1,\varphi_2,\dots$ into a shared ride. In the following, we consider the adoption dynamics of $N$ concurrent ride-hailing users with homogeneous preferences $a$ and $b$ in such a simplified one-to-many demand setting with a set of equidistant destinations on a ring with radius $R$ around the origin $o$ (see Fig.~\ref{fig:FIG1_model}b,i). In this setting, the direct trip distance is identical for all destinations, $d_\mathrm{req} = R$ and we identify the possible destinations by the angle $\varphi \in [0, 2\pi)$ in polar coordinates. We randomly select the destinations of the $N$ users independently and uniformly from all destinations and further assume that at most two users are matched into a shared ride. 
If two users with destinations $\varphi_1$ and $\varphi_2$ are matched into a shared ride, the user dropped off second experiences a detour given by the direct geometric distance between the two destinations (see Fig.~\ref{fig:FIG1_model}b,ii).
This detour depends only on the difference $\varphi_2 - \varphi_1$ and can be expressed via the law of cosines as
\begin{equation}
    d_\mathrm{det}(\varphi_1, \varphi_2) = R\,\sqrt{2 - 2\,\cos\left(\varphi_2 - \varphi_1\right)} \,. \label{eq:detour}
\end{equation}
The service provider matches shared ride requests to minimize the total distance driven including the return trip to the origin. With this matching strategy the provider also naturally minimizes the cumulative detour for all users. However, the provider will match two shared ride requests regardless of their destination on the ring (i.e. also if they are on opposite sides) since one shared trip is always shorter or at most equal in length compared to two single trips including their return trips.

\subsection{Replicator dynamics}
In such a basic one-to-many setting, the adoption of ride-sharing is characterized by the probability $p(\varphi,t) \in [0,1]$ of the sub-population of users with destination $\varphi$ to request a shared ride. We model the evolution of this sharing adoption with time-continuous replicator dynamics \cite{Cressman2014} 
\begin{equation}
    \frac{\partial p(\varphi,t)}{\partial t} = p(\varphi,t)\,\left(1-p(\varphi,t)\right)\, E\left[\Delta u(\varphi,t)\right] \,. \label{eq:replicator_dynamics}
\end{equation}
If the average utility of a shared ride is larger than the utility of a single ride, $E\left[\Delta u(\varphi,t)\right] > 0$ (see Fig.~\ref{fig:FIG1_model}a,i), users increase their sharing adoption $p(\varphi,t)$ over time. If the utility of a shared ride is smaller, $E\left[\Delta u(\varphi,t)\right] < 0$ (see Fig.~\ref{fig:FIG1_model}a,ii), they decrease it.

These dynamics may be interpreted in terms of the sub-population of users with destination $\varphi$ changing their overall adoption probability. Each realization of the game corresponds to a small time window in which concurrent requests are matched. During one day, a large number of different users with the same destination $\varphi$ make a trip and update their individual adoption decisions, resulting in the overall chage of the adoption probability of the whole sub-population as described by replicator dynamics. Here, we assume that the external conditions such as pricing and the overall demand distribution change on a much slower time scales than the adoption dynamics. We thus concentrate on the emerging stationary states of Eq.~\eqref{eq:replicator_dynamics}, where no user group (identified by their destination $\varphi$) can increase their utility by changing their decision unilaterally.

The absolute magnitude of the utility differences $\Delta u$ determines the time scale of the dynamics in Eq.~\eqref{eq:replicator_dynamics} but does not change the stationary states. The stationary sharing adoption only depends on the relative importance $\beta = b/a$ of detour compared to financial incentives. In the following we thus set $a=1$ and $R=1$ without loss of generality. The expected utility difference Eq.~\eqref{eq:expected_utility} becomes 
\begin{equation}
    E\left[\Delta u(\varphi,t)\right] = 1 - \beta \, E\left[d_\mathrm{det}(\varphi, t)\,\middle|\,\mathrm{share}\right] \, \label{eq:utility_general}
\end{equation}
with a single free parameter $\beta$.

\subsection{Simulation approach}
To simulate the dynamics, we discretize the set of destinations $\varphi \in \left[0, 2\pi\right)$ into $360$ distinct destinations. In each time step, we compute estimates for $E\left[\Delta u(\varphi,t)\right]$ by simulating the ride-sharing decisions, the matching of shared ride-requests, and the resulting utilities over $1000$ realizations per destination. For each realization, we fix one ride-hailing user $j=1$ who requests a shared ride to destination $\phi_1 = \varphi$ and select the destinations $\phi_j$, $j \in \{2,\dots \,N\}$, of the remaining $N-1$ users independently and uniformly randomly from all destinations. Each user realizes their sharing decision according to their current adoption probability (see Fig.~\ref{fig:FIG1_model}b,i), requesting a shared ride with probability $p(\phi_j, t)$ and a single ride with probability $1-p(\phi_j, t)$. We then match the shared ride requests to minimize the total distance driven, selecting uniformly between identical options (e.g. when three users request a shared ride to the same destination, Fig.~\ref{fig:FIG1_model}b,ii). We compute the utility (see Fig.~\ref{fig:FIG1_model}b,iii) for the focal user $j=1$ and estimate the expected utility of a shared ride and thereby the expected utility difference $E\left[\Delta u(\varphi,t)\right]$ as the average over all $1000$ realizations. After computing the expected utility for all destinations $\varphi$, we advance the sharing adoption by one Euler-step $\Delta t$ following Eq.~\eqref{eq:replicator_dynamics} and repeat the process. In this numerical evaluation, we restrict $p(\varphi,t)$ to values between $10^{-3}$ and $1-10^{-3}$ to avoid extremely long transients when the system gets close to the trivial fixed points $p(\varphi,t) = 0$ or $p(\varphi,t) = 1$. Due to the complex evaluation for each update of the replicator dynamics, we choose a default time step $\Delta t = 1$ for all simulation unless explicitly stated otherwise.

\section{Results}
\subsection{Two concurrent users}
We first consider the simplest setting with exactly $N=2$ concurrent users who request a ride at the same time, resulting in a simple matching problem. The users are matched into a shared ride if and only if both users request a shared ride. Otherwise, a user making a shared ride request receives the financial discount but is served in a single ride without incurring any detour. With the detour Eq.~\eqref{eq:detour}, we explicitly write the expected value of the utility difference $E\left[\Delta u(\varphi,t)\right]$ between a shared and a single ride for a user with destination $\varphi$ as the integral over all destinations $\phi$ of the other user with the corresponding detour if they also request a shared ride with probability $p(\phi,t)$,
\begin{eqnarray}
    \frac{\partial p(\varphi,t)}{\partial t} &=& p(\varphi,t)\,\left(1-p(\varphi,t)\right) \label{eq:two_user_replicator_dynamics}\\
    &\times&  \frac{1}{2\pi}\,\int_0^{2\pi} \left[1 - p(\phi,t) \, \frac{\beta \sqrt{2 - 2\,\cos\left(\varphi - \phi\right)}}{2} \right] \, \mathrm{d}\phi \,. \nonumber 
\end{eqnarray}
The factor $1/2$ in the detour term captures the expected value given that only one (randomly chosen) of the two users in a shared ride experiences a detour while the other is driven directly to their destination (compare also Fig.~\ref{fig:FIG1_model}b,ii).

\begin{figure}[t]
    \centering
	\includegraphics[width=\columnwidth]{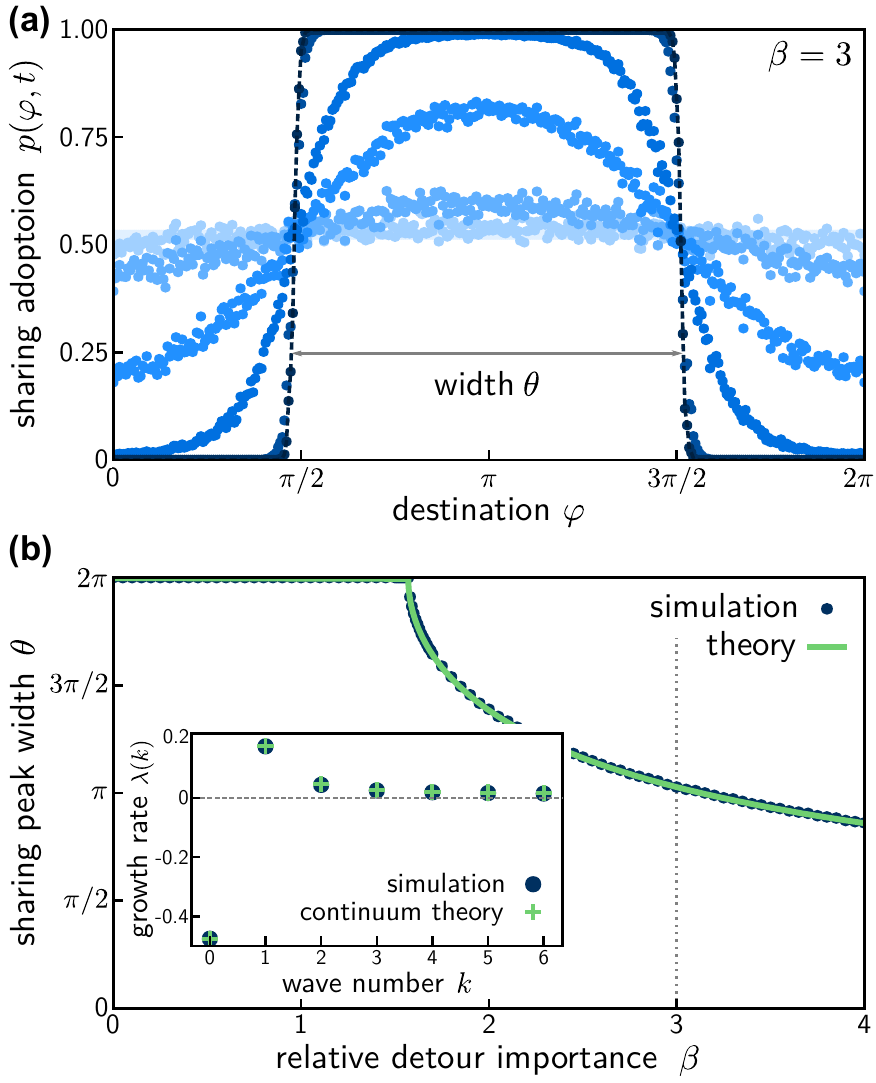}
    \caption{
        \textbf{Symmetry breaking in ride-sharing adoption dynamics.}  
        (a) The sharing adoption $p(\varphi, t)$ with $N=2$ concurrent users evolves from an initially homogeneous sharing adoption [Eq.~\eqref{eq:two_user_steady_state}, light blue] towards a stable single-peak sharing adoption (dark blue) where some users always request a shared ride while others never do.
        The plot shows the sharing adoption $p(\varphi,t)$ at times $t \in \{0,20,30,40,50,100,150\}$ (from light to dark) with a relative detour importance $\beta = 3$. The sharing adoption evolves following the replicator dynamics [Eq.~\eqref{eq:replicator_dynamics}] with a time step $\Delta t = 0.05$. We center the sharing peak at $\varphi = \pi$ for a clearer visualization. 
        (b) Stable steady states characterized by the width of the single-peak adoption pattern. Data points are created by slowly varying the relative detour importance $\beta$ in steps of $0.05$ ($0.01$ close to the transition to full sharing), letting the system settle into the equilibrium adoption state for $T=1000$ time steps after each change, and tracking the width of the sharing peak in the steady state. 
        (inset) A linear stability analysis of the replicator dynamics [Eq.~\eqref{eq:two_user_replicator_dynamics}, see appendix for details] with relative detour importance $\beta = 3$ predicts a single peak as the most unstable mode, consistent with the observations from direct numerical simulations (panel a). Crosses denote results of the analytical stability analysis with a continuous destination space. Circles denote results from the exact evaluation in discretized destination space. 
    }
    \label{fig:FIG2_pattern}
\end{figure}

Solving the two-user replicator dynamics, Eq.~\eqref{eq:two_user_replicator_dynamics}, for a homogeneous steady state $p(\varphi, t) = p^*$ such that $\left.\frac{\partial p(\varphi,t)}{\partial t}\right|_{p^*} = 0$ we find
\begin{equation}
    p^* = \frac{\pi}{2\beta} \,. \label{eq:two_user_steady_state}
\end{equation}
We provide the full calculation in the appendix. As expected, the homogeneous steady state sharing adoption increases as the relative detour importance $\beta$ decreases. However, simulating the replicator dynamics from this steady state (Fig.~\ref{fig:FIG2_pattern}a) reveals that this state is indeed unstable. The symmetry in the system is broken by random fluctuations in the estimation of the expected utility differences. Over time, the ride-sharing adoption increases on one side of the ring and decreases on the other until finally a single-peak sharing-nonsharing-pattern emerges. Figure~\ref{fig:FIG2_pattern}b illustrates how the total sharing adoption, characterized by the width $\theta$ of the sharing peak, changes with the relative detour importance $\beta$. While the total sharing adoption still increases as the relative detour importance decreases, it does so with a spatially heterogeneous pattern until the relative detour importance decreases below a critical value $\beta_c = \pi / 2$ where $p^* = 1$.

A stability analysis of the two-user replicator dynamics confirms this observation (Fig.~\ref{fig:FIG2_pattern}b inset). While the homogeneous steady state is stable with respect to spatially homogeneous changes of the sharing adoption, it is unstable to spatially dependent perturbations. In particular, the most unstable mode with the largest growth rate $\lambda(k) > 0$ is a single-period sinusoidal perturbation, $p(\varphi,t) = p^* + \delta \, \cos\left( k\,\varphi \right) \, e^{\lambda(k)\,t} + \mathcal{O}\left(\delta^2\right)$ with wavenumber $k = 1$, increasing the adoption on one side of the ring and decreasing it on the opposite side.

\subsection{Response function of ride-sharing adoption}
To understand the mechanism behind this symmetry breaking, we quantify the impact of a change in adoption of users with destination $\phi$ on other users. We consider a small change $\epsilon$ of the adoption probability localized to a small range of destinations $[\phi,\phi + \delta \phi]$ around a reference state $p_0(\varphi)$,
\begin{eqnarray}
    p(\varphi) &=& p_0(\varphi) + \delta p(\varphi; \epsilon, \phi) \\
    &=& p_0(\varphi) + \epsilon \, \Theta(\varphi - \phi) \, \Theta(\phi + \delta \phi - \varphi) \,,\nonumber
\end{eqnarray}
where $\Theta(\cdot)$ denotes the Heaviside-function. We now define the response function
\begin{equation}
    \chi_{p_0(\varphi)}(\varphi, \phi) = \lim_{\epsilon \rightarrow 0} \lim_{\delta \phi \rightarrow 0} \frac{ E\left[\Delta u(\varphi)\right]_{p(\varphi)} - E\left[\Delta u(\varphi)\right]_{p_0(\varphi)} }{\epsilon\,\delta \phi} \,, \label{eq:response_function}
\end{equation}
measuring the change in expected utility for users with destination $\varphi$ as a result of the change in adoption probability of users with destination $\phi$. If the response function $\chi(\varphi, \phi)$ is positive, the users with destination $\varphi$ would be more likely to share rides compared to the reference state. If it is negative, they would be less likely to share. 

In the two-user setting, we directly evalute the response function by 
expanding the expected utility [last term in Eq.~\ref{eq:replicator_dynamics}] in terms of $\epsilon$ and $\delta \phi$ such that the response function is equal to the contribution of destination $\phi$ to the expected detour,
\begin{equation}
    \chi(\varphi, \phi) = \frac{\beta}{2\pi}\,\frac{\sqrt{2 - 2\,\cos\left(\varphi - \phi\right)}}{2} \,. \label{eq:two_user_response}
\end{equation}
We also evaluate the response function numerically by comparing the expected detour in the homogeneous steady state $p^*$ and increasing the sharing adoption of a single destination $\phi$ by $\epsilon = 0.01$.

\begin{figure}[ht]
    \centering
	\includegraphics{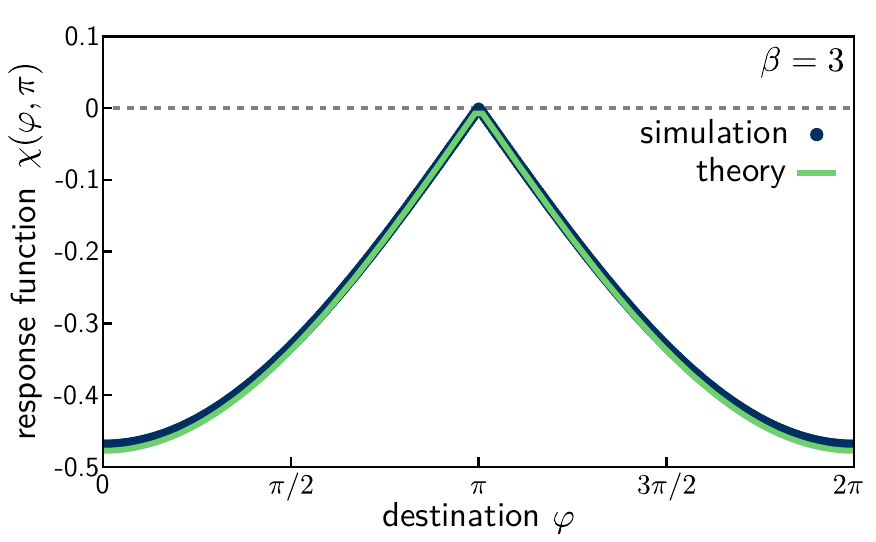}
    \caption{\textbf{Effective interaction with long-range inhibition explains symmetry breaking.}  
        Response function $\chi(\varphi, \pi)$ [Eq.~\ref{eq:response_function}] for $N=2$ users with relative detour importance $\beta = 3$ when users with destination $\phi = \pi$ are more likely to request a shared ride, evaluated around the homogeneous steady state $p^*(\varphi) = \pi/(2\beta)$. The solid green line shows the result of the direct evaluation of the response function in continuous destination space [Eq.~\ref{eq:two_user_response}]. The circles show the result of the numerical evaluation in discretized destination space with an increase $\epsilon = 0.01$ of the sharing probability at destination $\phi$. Far away destinations ($\varphi = 0$ or $2\pi$) are disincentivized to share by larger expected detours. Close by destinations around $\phi = \pi$ are less affected. The interactions between the adoption decisions of users with different destinations exhibits long range inhibition, explaining the emergence of a localized sharing peak.
    }
    \label{fig:FIG3_effectiveInteraction}
\end{figure}

Figure~\ref{fig:FIG3_effectiveInteraction} illustrates the response to an increase of the sharing adoption of users with destination $\phi = \pi$ in the center of the plot. The expected utility of shared rides decreases for all user. This effect is small for users with destinations close to $\phi = \pi$ as their detour is small when matched in a shared ride with a user with destination $\phi$. For destinations on the opposite side of the circle ($\varphi = 0$ or $\varphi = 2\pi$), these detours are large and the effect is stongest. 

Effectively, increasing the sharing adoption in one location strongly increases the expected detour for users with far-away destinations while the expected detour for close-by destinations is only weakly affected. This long-range inhibition in the detour-mediated interactions of the users promotes the formation of a localized single-peak pattern where users with certain destinations always request shared rides while users with other destinations never request shared rides.

\subsection{$N$ concurrent users}
\begin{figure}[t]
    \centering
	\includegraphics[width=\columnwidth]{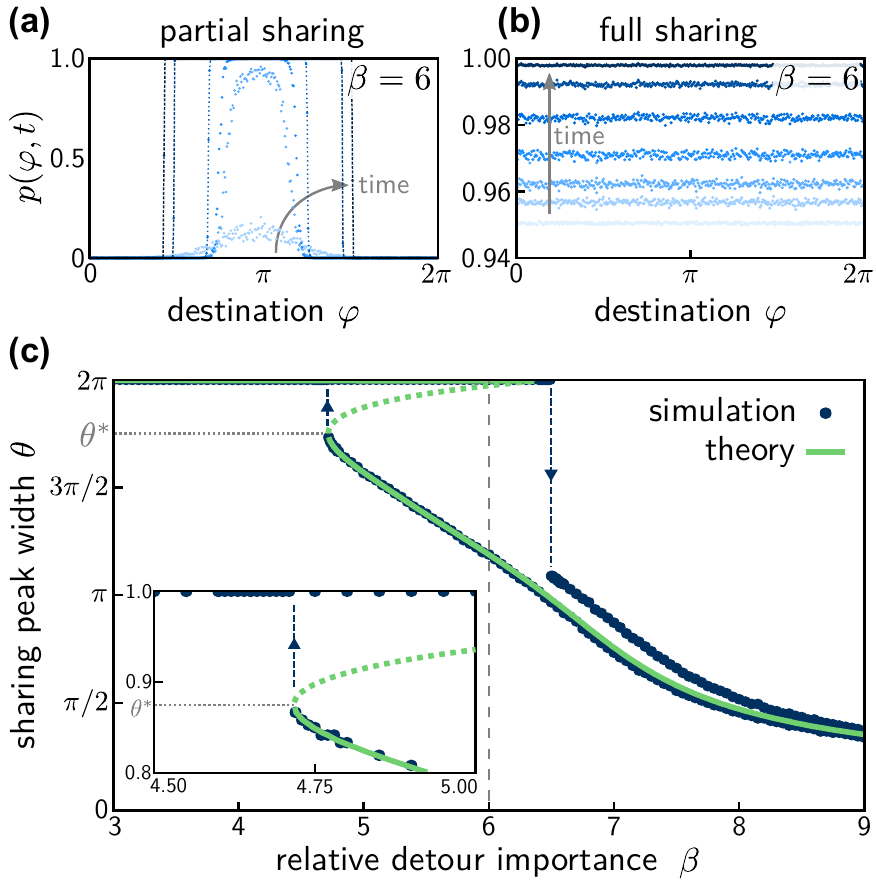}
    \caption{
        \textbf{Bistability of ride-sharing adoption.}  
        (a,b) Time evolution (light to dark) of the sharing adoption from different initial conditions for $N = 16$ concurrent users. Starting from low sharing adoption, $p(\varphi,0) \in \left[10^{-3}, 2\times 10^{-3}\right]$ uniformly randomly, a single-peak sharing pattern evolves and the system slowly converges to a stable peak width $\theta < 2\pi$ (panel a, $t \in \{0,20,30,40,100,1000,2000\}$ with simulation time step $\Delta t = 0.1$). For the same parameters, the full sharing state is stable and the system quickly returns to full sharing when initial adoption is high and homogeneous, $p(\varphi,0) \in \left[0.950, 0.951\right]$ uniformly randomly (panel b, $t \in \{0,1,2,4,8,16,32\}$ with simulation time step $\Delta t = 0.1$).
        (c) Bifurcation diagram of the stable sharing patterns (compare Fig.~\ref{fig:FIG2_pattern}). Ride-sharing adoption evolves to full adoption (peak width $\theta = 2\pi$) for low detour importance, $\beta < \beta_{c,\mathrm{part}}$. When the relative detour importance $\beta$ increases, the full sharing state becomes unstable above a critical value $\beta_{c,\mathrm{full}}$. However, the partial sharing state becomes stable in a saddle-node bifurcation already at $\beta_{c,\mathrm{part}} < \beta_{c,\mathrm{full}}$ and both adoption states coexist for intermediate values of $\beta$. The dashed and solid line (theory, light green) denotes the theoretical prediction from the stability condition [Eq.~\eqref{eq:N_user_stability_condition}], including an unstable partial sharing state (top branch, dashed) for $\beta_{c,\mathrm{part}} < \beta < \beta_{c,\mathrm{full}}$. In contrast to continuous destination space, the stability conditions for a growing and shrinking sharing peak do not exactly coincide due to self-interactions of users with the same destination (i.e. a non-zero probability of two users going to the same destination in discrete destination space), resulting in the different peak widths in the partial sharing state for large relative detour importance.
    }
    \label{fig:FIG3_bistability}
\end{figure}

With more than two users, the qualitative interactions between users and their sharing adoption $p(\varphi,t)$ at different locations $\varphi$ remains robust. Figure~\ref{fig:FIG3_bistability}a illustrates the emergence of a ride-sharing adoption pattern with a single-peak for $N = 16$ concurrent users starting from a homogeneous state with low sharing adoption. Interestingly, Fig.~\ref{fig:FIG3_bistability}b suggests that, in the same setting, the full sharing state $p(\varphi,t) = 1$ is stable as well. An initially high, homogeneous sharing adoption evolves to this full sharing state instead of breaking down into a single-peak sharing pattern. A single user outside an existing sharing peak does not want to share due to a large expected detour as no other users are requesting a shared ride with a similar destination. At the same time, however, if all users adopt ride-sharing, the expected detours become small due to the large number of users (i.e. users are more likely matched with another user with a close-by destination). 

This bistability is also directly predicted by the expected utility: We assume that the sharing adoption $p^*_\theta$ takes the form of a single sharing peak with width $\theta$, $p^*_\theta(\varphi) = 1$ if $0 \le \varphi \le \theta$ and $p^*_\theta(\varphi) = 0$ otherwise. The stability of such an adoption pattern is determined by the decisions of users on the edge of the sharing peak given by the expected utility difference $E\left[\Delta u(\theta)\right]_{p^*_\theta}$. Unfortunately, a direct analytical calculation would involve a large number of case distinctions to discern which users are actually matched. Here we numerically evaluate the expected detour over $10^8$ realizations of the destinations and sharing decisions of the other users to compute the expected utility Eq.~\eqref{eq:utility_general}.

In continuous destination space, the sharing peak width remains unchanged if the user on the edge is indifferent between shared and single rides, i.e. if the utility difference
\begin{equation}
    E\left[\Delta u(\theta)\right]_{p^*_\theta} = 1 - \beta\,E\left[d_\mathrm{det}(\theta)\,\middle|\,\mathrm{share}\right]_{p^*_\theta} = 0 \,, \label{eq:N_user_stability_condition_zero}
\end{equation}
giving the critical detour importance
\begin{equation}
     \beta_c(\theta) = \frac{1}{E\left[d_\mathrm{det}(\theta)\,\middle|\,\mathrm{share}\right]_{p^*_\theta}} \,. \label{eq:N_user_stability_condition}
\end{equation}
This critical detour importance also predicts where the full sharing state with $\theta  = 2\pi$ becomes unstable. In discrete destination space, we compute the expected detour for a user directly outside the sharing peak and thus find the critical detour importance where the width of the sharing peak would not grow any further. This correctly predicts the width of the sharing peak as we slowly decrease the relative detour importance $\beta$, always letting the system settle into its stable sharing adoption state. However, due to subtle differences in a discrete destination setting (the probability for two users to go to the same destination is non-zero), it is slightly different from the stability condition for an existing sharing peak breaking down when we slowly increase the relative detour importance. 

\begin{figure}[t]
    \centering
	\includegraphics[width=\columnwidth]{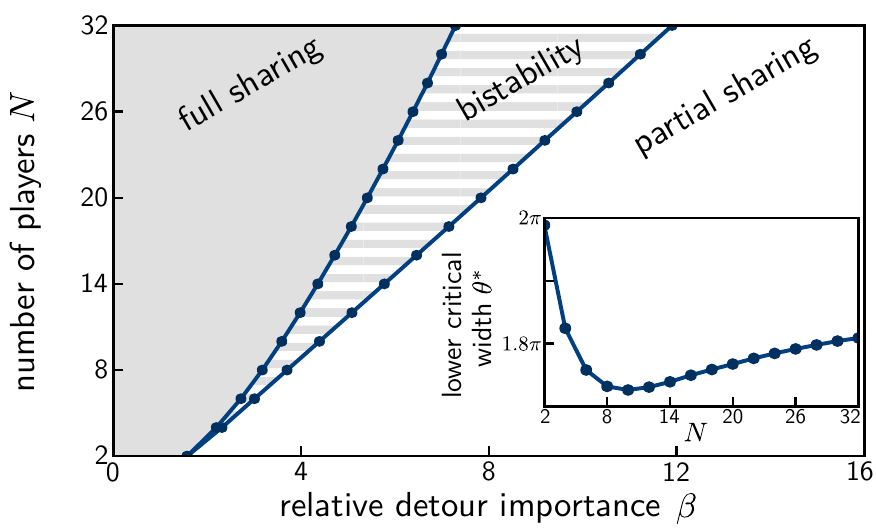}
    \caption{
        \textbf{Bistability of ride-sharing adoption increases in high-demand settings.}
        Bifurcation diagram of ride-sharing adoption as a function of the detour importance and the number of concurrent users (compare Fig.~\ref{fig:FIG3_bistability}c for $N = 16$ users). The left points indicate the lower critical detour importance $\beta_{c,\mathrm{part}}$ where the saddle-node bifurcation of the partial sharing state $\theta < 2\pi$ occurs. The right points indicate the upper critical detour importance $\beta_{c,\mathrm{full}}$ where the full sharing state $\theta = 2\pi$ becomes unstable. 
        We compute the critical detour importance based on the stability condition Eq.~\eqref{eq:N_user_stability_condition} together with numerical evaluation of the expected detours for users with destination immediately outside the sharing peak with a given width $\theta$ over $10^8$ realizations. (inset) Width $\theta^*$ of the sharing peak at the lower critical point $\beta_{c,\mathrm{part}}$. For intermediate numbers of users this lower critical width is minimal, indicating a large loss of efficiency compared to the full sharing state. For more users, the range of the bistability increases together with the lower critical width and the system becomes more efficient even if it is not in the full sharing state.  
    }
    \label{fig:FIG4_bifurcation}
\end{figure}

Figure~\ref{fig:FIG3_bistability}c shows the bifurcation diagram for $N = 16$ users, illustrating the bistability of the partial ($\theta < 2\pi$) and full ($\theta = 2\pi$) sharing state as well as the saddle-node bifurcation where the partial sharing state disappears for decreasing relative detour importance $\beta$. For small peak widths $\theta < \theta^*$ the sharing peak is always stable. Due to the uniform request distribution, users with destinations inside the sharing peak always experience shorter expected detours (and thus prefer sharing) while users further outside always experience longer detours (and thus prefer single rides). However, for sufficiently large peak widths $\theta > \theta^*$, detours become smaller as the peak width increases since users are matched around the circle. The sharing peak is unstable.

The critical detour importance increases as the number of users increases (and the expected detour decreases), making it easier to achieve full sharing adoption. However, at the same time, the width of the bistability region also increases with the number of users. Figure~\ref{fig:FIG4_bifurcation} shows the full bifurcation diagram as a function of the relative detour importance and the number of concurrent users. Intriguingly, the width $\theta^*$ of the sharing peak where the partial sharing regime undergoes the saddle-node bifurcation is minimal for intermediate numbers of ride-hailing users (see Fig.~\ref{fig:FIG4_bifurcation} inset). This suggests a potentially larger loss of ride-sharing adoption in these intermediate-demand settings compared to the extreme cases of low or high demand.

\section{Discussion}
The basic mechanism underlying the bistability between partial and full adoption observed in our model is similar to the bistability in classic models of technology adoption with network effects \cite{Easley2010, Schroder2018, Arthur1996, Katz1994}, i.e.~if the utility of the technology or service relies on its wide-spread adoption, like for example for communication technologies or online social platforms. However, the inherent spatial component of ride-sharing services and the strategic interactions between users with different destinations give rise to symmetry breaking dynamics and spatially heterogeneous adoption patterns. The bistability between homogeneous and heterogeneous adoption patterns may lead to a loss in efficiency of ride-sharing services due to lower than expected adoption. This observation is particularly relevant for emerging services with a small initial user base and low sharing adoption, revealing a potential barrier to achieving full sharing adoption solely due to the collective interactions among users. Similar effects may partially underlie currently low ride-sharing adoption in metropolitan areas such as New York City \cite{Storch2020}.

We have studied the dynamics of ride-sharing adoption in a highly simplified setting. However, our observation and the qualitative, fundamental interactions between users are likely persistent in more complex settings as well. For example, potential compensations schemes \cite{furuhata2014online} to equalize the utility for both users in a shared ride are equivalent to the dynamics in our model system. Since users are equally like to pay the compensation as they are to receive it, such schemes cannot change the expected utility of a share ride. Other modifications such as limiting the detour when matching rides \cite{kucharski2020exact} may increase the adoption of sharing but at the same time would reduce the number of actually matched rides. Despite this robustness, in real-world ride-sharing applications, additional influences from alternative mode choice options, matching strategies, or quickly evolving regulatory boundary conditions add to the dynamics \cite{Alonso-Gonzalez2020, Morris2019, Sarriera2017, LO2018, Schwieterman2019}. In particular, adoption patterns in real ride-sharing applications seem to be dominated by socio-economic heterogeneities and the interaction with other (public) transport modes \cite{Storch2020, Schwieterman2019}. While this observation suggests that the pattern formation dynamics found in the simplified model may not be dominant, the spatially heterogeneous willingness to share may still result in higher difficulty to achieve large-scale ride-sharing adoption due to the inherent bistability of partial ride-sharing adoption patterns and full sharing adoption. Moreover, additional disincentives from inconvenience and loss of privacy during a shared ride may further limit the adoption of ride-sharing.

More complex urban settings such as two-dimensional street networks, heterogeneous demand distributions with many-to-many routing, and more complex matching schemes with on-route pickups may give rise to more complex interactions between the users and to more intricate adoption patterns. Unfortunately, the complexity of evaluating the expected utility and the dependence on the realization of the matching of shared rides renders the (numerical) evaluation of the dynamics a non-trivial problem already in our simplified system. 

Overall, our results demonstrate the possibility of spontaneous symmetry breaking and self-organized pattern formation dynamics in shared urban mobility. Importantly, these complex dynamics need not be rooted in the algorithms underlying new types of mobility services, spatially distributed vehicle fleets, or inherent heterogeneities, but may ultimately result from interactions among the users alone. These findings emphasize the importance of understanding the collective dynamics of both the supply \cite{Schroeder2020} and demand sides \cite{Storch2020} of ride-sharing, and the interactions between them, to enable more sustainable mobility \cite{Sovacool2020}. Revealing the fundamental detour-mediated interactions between the users as the cause of symmetry breaking and pattern formation dynamics of ride-sharing adoption, our modeling approach may contribute to a better understanding of the (non)adoption of ride-sharing services. This may ultimately help to design services and incentives that promote the adoption of ride-sharing and other forms of sustainable shared mobility.\\

\section*{Acknowledgements}
We thank Nora Molkenthin for helpful discussions. D.S. acknowledges support from the Studienstiftung des Deutschen Volkes. M.T. acknowledges support from the German Research Foundation (Deutsche Forschungsgemeinschaft, DFG) through the Center for Advancing Electronics Dresden (cfaed).

\bibliography{lit}

\cleardoublepage

\appendix

\renewcommand{\thefigure}{A\arabic{figure}}
\renewcommand{\thetable}{A\arabic{table}}
\renewcommand{\theequation}{A\arabic{equation}}

\setcounter{figure}{0}
\setcounter{table}{0}
\setcounter{equation}{0}

\onecolumngrid

\section*{Appendix}

\subsection*{$N=2$ user model with continuous destinations}
The full replicator equation for $N=2$ users is given by Eq.~\eqref{eq:two_user_replicator_dynamics} in the main text
\begin{equation}
    \frac{\partial p(\varphi,t)}{\partial t} = p(\varphi,t)\,\left(1-p(\varphi,t)\right)
    \times \frac{1}{2\pi} \, \int_0^{2\pi} \left[1 - p(\phi,t) \, \frac{\beta \sqrt{2 - 2\,\cos\left(\varphi - \phi\right)}}{2} \right] \, \mathrm{d}\phi \,.
    \label{eq:two_user_replicator_equation_appendix}
\end{equation}

\subsubsection*{Homogeneous steady state}
We look for homogeneous steady states $p(\varphi,t) = p^*=\mathrm{const.}$ defined by $\left.\frac{\partial p(\varphi,t)}{\partial t}\right|_{p^*} = 0$. From the structure of Eq.~\eqref{eq:two_user_replicator_equation_appendix}, we quickly find two trivial fixed points at $p^*=0$ and $p^*=1$. These fixed points are the result of the mathematical structure of the replicator equation (if one strategy dies out completely, it cannot reproduce). We find another fixed point $0 < p^* < 1$ by setting the integral in Eq.~\eqref{eq:two_user_replicator_equation_appendix} to zero:
\begin{equation}
    1 - \frac{\sqrt{2}\,\beta\, p^*}{4\pi} \int_0^{2\pi} \sqrt{1 - \cos\left(\varphi - \phi\right)}\, \mathrm{d}\phi \stackrel{!}{=} 0 \,.
    \label{eq:two_user_fixed_point_appendix}
\end{equation}
The remaining integral 
can be solved by applying the half-angle identity $\sin^2(x/2) = \frac{1 - \cos(x)}{2}$ such that the integrand becomes $\sqrt{1-\cos \left( \varphi-\phi\right) } = \sqrt{2}\,\left|\sin\left(\frac{\varphi-\phi}{2}\right)\right|$.
Using the $2\pi$ periodicity and the symmetry of our solution, we set $\varphi = 0$ without loss of generality. The value of $\sin\left(\frac{-\phi}{2}\right)$ is negative over the whole integration range. We use the antisymmetry of the sine-function to replace the absolute value by changing the sign of the argument and compute the integral as
\begin{equation}
    \int_{0}^{2\pi}\sqrt{1-\cos\left(\varphi-\phi\right) }\,\mathrm{d}\phi = \sqrt{2}\int_{0}^{2\pi}\left|\sin\left(\frac{\varphi-\phi}{2}\right)\right|\,\mathrm{d}\phi=\sqrt{2}\int_{0}^{2\pi}\sin\left(\frac{\phi}{2}\right)\mathrm{d}\phi=\sqrt{8}\,\int_{0}^{\pi}\sin\left(y\right)\mathrm{d}y=\sqrt{32} \,.
    \label{eq:two_user_integral_1_solution_appendix}
\end{equation}

Inserting this result in Eq.~\eqref{eq:two_user_fixed_point_appendix} leads to
\begin{equation}
        1 - \frac{\sqrt{2}\,\beta\, p^*}{4\pi} \sqrt{32} = 0 \,,
\end{equation}
which yields the fixed point
\begin{equation}
    p^* = \frac{\pi}{2\beta} \,, \label{eq:hom_steady_state_appendix}
\end{equation}
reproducing Eq.~\eqref{eq:two_user_steady_state} in the main text.

\subsubsection*{Linear stability analysis}

To analyze the linear stability of the two user system, we make the ansatz
\begin{equation}
p\left(\varphi,t\right)=p^*+\delta\cos\left(k\varphi\right)\mathrm{e}^{\lambda_{k}t} + \mathcal{O}\left(\delta^2\right)  
\end{equation}
of a small perturbation with wavenumber $ k\in \mathbb{N}$ around the homogeneous adoption state $p^*$ in Eq.~\eqref{eq:two_user_replicator_equation_appendix}. Inserting the ansatz gives
\begin{multline}
    \frac{\partial}{\partial t}\Big[p^*+\delta\cos(k\varphi)\mathrm{e}^{\lambda_{k}t} + \mathcal{O}\left(\delta^2\right) \Big] = \left(p^*+\delta\cos(k\varphi)\mathrm{e}^{\lambda_{k}t}+ \mathcal{O}\left(\delta^2\right)\right)\left(1-\left[p^*+\delta\cos(k\varphi)\mathrm{e}^{\lambda_{k}t}+ \mathcal{O}\left(\delta^2\right)\right]\right)\\
    \times \left[ 1 - \int_{0}^{2\pi} \frac{1}{2\pi}\cdot\left(p^*+\delta\cos\left(k\phi\right)\mathrm{e}^{\lambda_{k}t}+ \mathcal{O}\left(\delta^2\right)\right)\left(\frac{\beta}{2}\sqrt{2-2\cos\left( \varphi-\phi\right) }\right)d\phi\right]\,. \label{eq:two_user_stability_analysis_start_appendix}
\end{multline}

The evaluation of the last part of Eq.~\eqref{eq:two_user_stability_analysis_start_appendix} leads to two integrals:
\begin{multline}
        1 - \int_{0}^{2\pi} \frac{1}{2\pi}\cdot\left(p^*+\delta\cos\left(k\phi\right)\mathrm{e}^{\lambda_{k}t}\right)\left(\frac{\beta}{2}\sqrt{2-2\cos\left( \varphi-\phi\right) }\right)\mathrm{d}\phi\\
        =1 
        -\frac{p^*}{2\pi}\left[\frac{\sqrt{2}\beta}{2}\int_{0}^{2\pi}\sqrt{1-\cos\left( \varphi-\phi\right) }\,\mathrm{d}\phi\right]
        -\frac{\delta\mathrm{e}^{\lambda_{k}t}}{2\pi}\left[\frac{\sqrt{2}\beta}{2}\int_{0}^{2\pi}\cos\left(k\phi\right)\sqrt{1-\cos\left( \varphi-\phi\right) }\,\mathrm{d}\phi\right]\,.
        \label{eq:two_user_stability_analysis_second_bracket}
\end{multline}
The first integral is already known from Eq.~\eqref{eq:two_user_integral_1_solution_appendix} and cancels with the  $1$ when evaluated at the homogeneous steady state Eq.~\eqref{eq:hom_steady_state_appendix}. To evaluate the second integral, we repeatedly apply integration by parts, assuming the integrand to be of the shape $f^\prime\cdot g$ with $f^\prime=\cos\left(k\varphi\right)$ and $g=\sqrt{1-\cos\left( \varphi-\phi\right) }$, which leads to
\begin{equation}
    \int_{0}^{2\pi}f^\prime(\varphi)\, g(\varphi,\phi)\,\mathrm{d}\phi=\left.\frac{1}{k}\sin\left(k\phi\right)\sqrt{1-\cos\left( \varphi-\phi\right) }\right|_{0}^{2\pi}-\frac{1}{k}\int_{0}^{2\pi}\sin\left(k\phi\right)\cdot\frac{1}{2}\frac{-\sin\left(\varphi-\phi\right)}{\sqrt{1-\cos\left( \varphi-\phi\right) }}\mathrm{d}\phi \,,
\end{equation}
where the first part vanishes when inserting the boundaries. To evaluate the remaining integral, we once again use the half angle identity $\sqrt{1-\cos \left( x\right) }=\sqrt{2}\left|\sin\left(\frac{x}{2}\right)\right|$ to simplify the expression and substitute $x=\frac{\varphi-\phi}{2}$ to find
\begin{equation}
    -\frac{1}{k}\int_{0}^{2\pi}\sin\left(k\phi\right)\frac{-\sin\left(\varphi-\phi\right)}{2\,\sqrt{1-\cos\left( \varphi-\phi\right) }} \, \mathrm{d}\phi = -\frac{1}{\sqrt{2}k}\int_{\frac{\varphi}{2}}^{\frac{\varphi}{2}-\pi}\sin\left(k\left(\varphi-2x\right)\right)\,\frac{\sin\left(2x\right)}{\left|\sin\left(x\right)\right|} \,\mathrm{d}x \,.
\end{equation}
With $\varphi\in\left[0,2\pi\right)$ the integration range is always in $\left[-\pi,\pi\right]$ for which the slightly modified double angle identity $\frac{\sin\left(2x\right)}{\left|\sin\left(x\right)\right|}=2\cos\left(x\right)\mathrm{sgn}\left(x\right)$ holds (with the exception of the isolated singular points $\{-\pi,0,\pi\}$). With this identity, we split the integral to resolve the explicit sign-function and the integral becomes
\begin{equation}
\frac{\sqrt{2}}{k}\int_{\frac{\varphi}{2}-\pi}^{\frac{\varphi}{2}}\sin\left(k\left(\varphi-2x\right)\right)\,\cos\left(x\right)\mathrm{sgn}\left(x\right) \, \mathrm{d}x = \frac{\sqrt{2}}{k}\left(\int_{0}^{\frac{\varphi}{2}}\sin\left(k\left(\varphi-2x\right)\right)\cdot\cos\left(x\right)\mathrm{d}x -
\int_{\frac{\varphi}{2}-\pi}^{0}\sin\left(k\left(\varphi-2x\right)\right)\cdot\cos\left(x\right)\mathrm{d}x\right) \,.
\label{eq:two_user_stability_analysis_step_positive_negative_appendix}
\end{equation}

These two integrals are of the same form, and have to be integrated by parts with $f^\prime(x)=\cos\left(x\right)$ and $g(\varphi,x)=\sin\left(k\left(\varphi-2x\right)\right)$:
\begin{equation}
\int_{a}^{b}\sin\left(k\left(\varphi-2x\right)\right)\cdot\cos\left(x\right)\mathrm{d}x=
\left[\sin\left(k\left(\varphi-2x\right)\right)\sin\left(x\right)\right]_{a}^{b}+2k\int_{a}^{b}\sin\left(x\right)\cos\left(k\left(\varphi-2x\right)\right)\mathrm{d}x \,.
\end{equation}
The resulting integral has to be integrated by parts once more, this time with the substitutions $f^\prime(x)=\sin\left(x\right)$ and $g(\varphi,x)=\cos\left(k\left(\varphi-2x\right)\right)$. We get

\begin{multline}
\int_{a}^{b}\sin\left(k\left(\varphi-2x\right)\right)\,\cos\left(x\right)\mathrm{d}x=\\
\left[\sin\left(k\left(\varphi-2x\right)\right)\sin\left(x\right)\right]_{a}^{b}+2k\left(\left[-\cos\left(x\right)\cos\left(k\left(\varphi-2x\right)\right)\right]_{a}^{b}+2k\int_{a}^{b}\cos\left(x\right)\sin\left(k\left(\varphi-2x\right)\right)\mathrm{d}x\right)\,,
\end{multline}
which we solve for the integral
\begin{equation}
\int_{a}^{b}\sin\left(k\left(\varphi-2x\right)\right)\,\cos\left(x\right)\mathrm{d}x=
\frac{\left[\sin\left(k\left(\varphi-2x\right)\right)\sin\left(x\right)\right]_{a}^{b}-2k\left[\cos\left(x\right)\cos\left(k\left(\varphi-2x\right)\right)\right]_{a}^{b}}{1-4k^2} \,.
\end{equation}
Plugging this result back into Eq.~\eqref{eq:two_user_stability_analysis_step_positive_negative_appendix} and evaluating the integrals with the corresponding boundaries give the result for the integral
\begin{equation}
    \frac{\sqrt{2}}{k}\left(\frac{-2k\left[\cos\left(\frac{\varphi}{2}\right)-\cos\left(k\varphi\right)\right]}{1-4k^2}
    -\frac{-2k\left[\cos\left(k\varphi\right)-\cos\left(\frac{\varphi}{2}-\pi\right)\right]}{1-4k^2}\right)=
    \frac{\sqrt{32}\,\cos\left(k\varphi\right)}{1-4k^2} \,.
\end{equation}
Inserting this result into Eq.~\eqref{eq:two_user_stability_analysis_start_appendix} yields 
\begin{multline}
    \frac{\partial }{\partial t}\Big[p^*+\delta\cos(k\varphi)\mathrm{e}^{\lambda_{k}t} + \mathcal{O}\left(\delta^2\right)\Big] 
    = \left(p^*+\delta\cos(k\varphi)\mathrm{e}^{\lambda_{k}t}+ \mathcal{O}\left(\delta^2\right)\right)\left(1-\left[p^*+\delta\cos(k\varphi)\mathrm{e}^{\lambda_{k}t}+ \mathcal{O}\left(\delta^2\right)\right]\right)\\
    \times \Bigg( 1 - \underbrace{\frac{p^*}{2\pi}\left[\frac{\sqrt{2}\beta}{2}\int_{0}^{2\pi}\sqrt{1-\cos\left( \varphi-\phi\right) }\,\mathrm{d}\phi\right]}_{=1\quad\mathrm{with\;Eq.\;(A3)\;and\;(A5)}}
        -\frac{\delta\mathrm{e}^{\lambda_{k}t}}{2\pi}\underbrace{\left[\frac{\sqrt{2}\beta}{2}\int_{0}^{2\pi}\cos\left(k\phi\right)\sqrt{1-\cos\left( \varphi-\phi\right) }\,\mathrm{d}\phi\right]}_{=\frac{4\beta\,\cos\left(k\varphi\right)}{1-4k^2} \quad\mathrm{with\;Eq.\;(A15)}} + \mathcal{O}\left(\delta^2\right)\Bigg) \,.
\end{multline}
The last factor is of order $\delta$ such that keeping only terms of order $\delta$ requires keeping only the constant terms $p^*\,\left(1-p^*\right)$ from the first two factors and finally gives the linearized equation around the homogeneous steady state 
\begin{equation}
    \delta\,\lambda_{k}\,\cos\left(k\varphi\right)\mathrm{e}^{\lambda_{k}t} + \mathcal{O}\left(\delta^2\right) = 
    \left(p^*-{p^*}^2\right)\,\left[ -\frac{\delta \, \mathrm{e}^{\lambda_{k}t}}{2\pi}\left[4\beta\frac{\cos\left(k\varphi\right)}{1-4k^2}\right] \right] + \mathcal{O}\left(\delta^2\right) \,.
\end{equation}
Comparing the linear order terms and solving for the growth $\lambda_k$ yields
\begin{equation}
    \lambda_{k}=\left(p^*-{p^*}^2\right)\,\left(-\frac{2\beta}{\pi}\frac{1}{1-4k^2}\right)= \frac{1-p^*}{4k^2-1} \,.
\end{equation}
The homogeneous steady state is stable with respect to spatially homogeneous perturbations, $\lambda_0<0$, but unstable for all inhomogeneous perturbation, $\lambda_k>0$ with wavenumber $k \ge 1$.

\subsubsection*{Derivation of the theoretical sharing peak width}
To predict the final steady state sharing pattern, we make an ansatz for the shape of the sharing adoption pattern based on the single peak pattern observed in the simulations. This sharing adoption pattern $p^*_\theta(\varphi)$ takes the form of a single sharing peak with width $\theta$, i.e. $p^*_\theta(\varphi) = 1$ if $0 \le \varphi \le \theta$ and $p^*_\theta(\varphi) = 0$ otherwise.
We now calculate the expected utility for an individual with a destination just at the border where the sharing regime and the non-sharing regime meet. As described in equation \ref{eq:N_user_stability_condition_zero} in the main text, a user with destination $\varphi = 0$ or $\varphi = \theta$ at the border between the sharing regimes should be indifferent between both options in the steady state, $\left.E[\Delta u(0)]\right|_{p_\theta^*} = \left.E[\Delta u(\theta)]\right|_{p_\theta^*} = 0$. Otherwise, the sharing probabilities at this point would either increase or decrease over time.

The sharing adoption
\begin{equation}
    p_\theta^*(\varphi) = \Theta(\theta-\varphi)
\end{equation}
with $\varphi \in [0, 2\pi)$ where $\Theta$ denotes the Heaviside function plugged into the expected utility integral yields
\begin{equation}
    \left.E[\Delta u(\theta)]\right|_{p_\theta}^*=\int_0^{2\pi} \frac{1}{2\pi} \left[1 - \Theta(\theta-\phi) \, \frac{\beta \sqrt{2 - 2\,\cos\left(\theta - \phi\right)}}{2} \right] \, \mathrm{d}\phi\overset{!}{=}0.
\end{equation}
The integral can be calculated by once again using the half-angle identity $\sqrt{1-\cos \left( x\right) }=\sqrt{2}\left|\sin\left(\frac{x}{2}\right)\right|$:
\begin{equation}
    1 - \frac{\sqrt{2}\beta}{4\pi}\int_0^{\theta} \sqrt{1 - \cos\left(\theta - \phi\right)} \, \mathrm{d}\phi=1 - \frac{\beta}{2\pi}\int_0^{\theta}\left|\sin\left(\frac{x}{2}\right)\right|\, \mathrm{d}x \,.
    \label{eq:two_user_theoretical_width_step_appendix}
\end{equation}
As $0\leq\theta\leq2\pi$, we neglect the absolute value and calculate the integral, resulting in:
\begin{equation}
    1 - \frac{\beta}{2\pi}\int_0^{\theta}\sin\left(\frac{x}{2}\right)\, \mathrm{d}x = 1- \frac{\beta}{\pi}\left(1-\cos\left(\frac{\theta}{2}\right)\right)\overset{!}{=}0
\end{equation}
which results in the equation 
\begin{equation}
    \cos\left(\frac{\theta}{2}\right) = 1-\frac{\pi}{\beta} \,.
\end{equation}
We thus find the theoretical prediction for the width of the sharing peak in a system with two concurrent users (compare Fig.~2b in the main text) 
\begin{equation}
    \theta = 2\arccos\left(1-\frac{\pi}{\beta}\right)
\end{equation}
as long as $\beta \geq \frac{\pi}{2}$. For $\beta = \frac{\pi}{2}$ we find $\theta = 2\pi$, i.e. the system is in the full sharing regime. For smaller values of the detour importance, we do not find a solution anymore since the utility difference $E\left[\Delta u(\varphi)\right]$ is always positive and there is no sharing peak. The system remains in the full sharing regime. We arrive at the final expression for the sharing regime width
\begin{equation}
    \theta = \begin{cases}
        2\arccos\left(1-\frac{\pi}{\beta}\right) & \beta \geq \frac{\pi}{2} \\
    2\pi & \mathrm{else}
    \end{cases}
\end{equation}

\end{document}